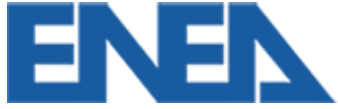

# Numerical investigation of a particle system compared with first and second gradient continua: Deformation and fracture phenomena*

**Antonio Battista**
*M & MoCS, International Research Center, Universitá dell'Aquila, Italy*

**Luigi Rosa**
*Physics Department, Universitá degli studi di Napoli Federico II, Italy*

**Ramiro dell'Erba**
*ENEA Technical Unit technologies for energy and industry – Robotics Laboratory*

**Leopoldo Greco**
*M & MoCS, International Research Center, Universitá dell'Aquila, Italy*

**Abstract**
A discrete system constituted of particles interacting by means of a centroid-based law is numerically investigated. The elements of the system move in the plane, and the range of the interaction can be varied from a more local form (first-neighbours interaction) up to a generalized $n$th order interaction. The aim of the model is to reproduce the behaviour of deformable bodies with standard (Cauchy model) or generalized (second gradient) deformation energy density. The numerical results suggest that the considered discrete system can effectively reproduce the behaviour of first and second gradient continua. Moreover, a fracture algorithm is introduced and some comparison between firstand second-neighbour simulations are provided.

## 1. Introduction

In the present paper a numerical investigation of the discrete mechanical system first described in [1] is performed. In particular, further comparison between the continuous case simulations (performed with the finite element method or FEM) and the proposed discrete system are provided. Moreover, the model described in [1] has been enriched to include fracture, and some numerical simulations with preliminary results are shown. The aim of the proposed model is to develop an appropriate numerical tool capable of modelling the behaviour of deformable bodies, and to take into account higher gradient constitutive relations (for interesting results in $n$th gradient theory the reader can see [2–17]; on the importance of this topic, see [18], where a general overview of recent results is provided).

It is worth noting that $n$th gradient theories can be contextualized in the more general framework of micromorphic/microstructured continua, which nowadays is a very active research field (see, for instance, [19–21] for classical references, [22–34] for interesting applications, [35–41] for recent theoretical results and [42] for a recent review). The main reason behind the interest in the aforementioned theoretical models lies in the fact that they have proven useful for the mathematical description of objects whose richness at the microscale cannot be captured by classical continuum models, that is, metamaterials (see, e.g., [43, 44] for reviews of recent results and [45–51] for interesting examples). The recent development of new techniques, such as three-dimensional (3D) printing or electro-spinning, gives the possibility to obtain increasingly complex and exotic micro-structures, which can provide a reasonably sound experimental basis. On the other hand (as often happens in the process of scientific research) the amount of new experimental data opens several deep and complex theoretical problems.

It is clear that in the presented background, numerical tools are essential in order to have a suitable mediation between theoretical and experimental results. In particular, in our opinion, a numerical investigation should be a good compromise between computational cost and accuracy of the results, as is required in rapid prototyping processes typical of modern technological research. Currently the well established methods of finite element analysis, and in particular isogeometric analysis are employed for studying deformation and fracture. The latter (see, e.g., [52–60]), can be especially convenient for shape optimization problems that easily arise in the study of multi-agent systems moving in unbounded domains and starting from arbitrary configurations, as is the case with our system. However, these numerical techniques are usually computationally expensive. The proposed algorithm offers the advantage of limited computational costs. Since the algorithm is based on a linear operation, that is, the computation of the centroid, its computational cost increases only linearly with the number of particles in the system. In [61], it is shown how the model may exhibit a rich range of behaviours, such as asymptotic convergence to the equilibrium configuration, instabilities of various kinds and, in well determined circumstances, spontaneous growth of the crack length after an almost steady state. The investigation of the instabilities of the system is not trivial, and represents a challenge for future researches (useful tools may be found in recent results developed in [62–64]).

## 2. The model

We start by briefly recalling the main features of the system.

The system is constituted of a finite number of particles that occupy, in the initial configuration, the nodes of a (squared) lattice with steps of 1 unit length (*l*), as graphically represented in Figure 1. In the same figure, first and second neighbours are also highlighted by means of a color map. Our concept of neighbourhood is based on the Chebyshev distance[1] between the elements in the initial configuration: two elements with, in $C^*$, a Chebyshev distance equal to $n$ are $n$th neighbours. Since the aim is to reproduce the characteristics of solid bodies, the concept of neighbours is Lagrangian, which means that neighbourhood is preserved during the time evolution of the system, with the only exceptions arising with the fracture algorithm (see Section 4). An element of the system, which we will call the *leader*, has an imposed motion defined for a discrete set of time steps $t_1, t_2, ... t_n ...$. The evolution of the system proceeds by setting, at each time step, an iterative process in which, at the $i$th step, the $i$th neighbours of the leader move to the centroid of their first neighbours. The process stops when all the elements of the system have moved, and then restarts at every following time step (for a more detailed description the reader is referred to [1]). In order to avoid spontaneous collapse of the edges (a wellknown edge effect in swarm robotics and other studies involving particle systems), an external frame constituted by 'fictitious' elements is introduced (see Figure 2). Every fictitious element translates rigidly with the closest 'true' element. The vertex true elements carry in rigid translation the vertex fictitious element as well as its two fictitious neighbours. In this way, every element has a full set of eight first neighbours for the computation of the centroid (see again [1] for the detailed description of the role of fictitious elements). The model exhibits pronounced nonlinear behaviours, as shown in [61], since composing motions for the leader does not lead to a simple superposition of effects in the configuration of the system.

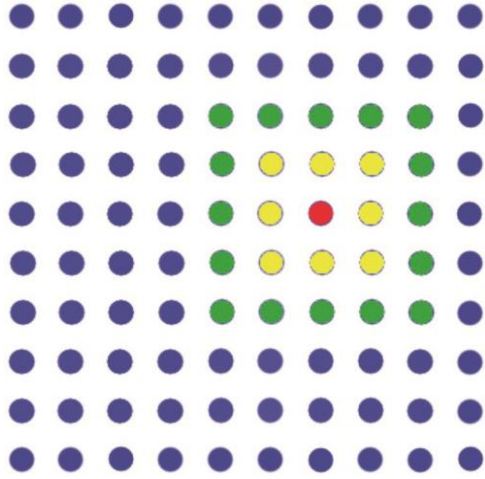

**Figure 1.** Graphical representation of neighbours. First and second neighbours of the red element are in yellow and green respectively.

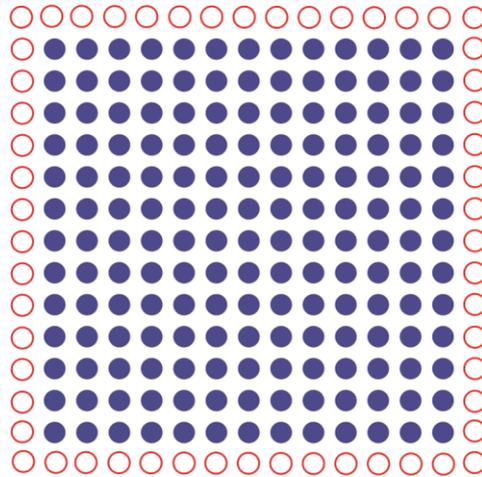

**Figure 2.** Graphical representation of fictitious elements. Every fictitious element (empty dots) translates rigidly with the closest true element. The vertex true elements carry in rigid translation the vertex fictitious element and its two fictitious neighbours.

The model can be straightforwardly generalized to $n$th order interaction if, in the above outlined algorithm, the computation of the centroid is performed over the set of $n$th neighbours of each element; in this case, an $n$th order fictitious boundary (i.e. $n$ external frames) has to be introduced as well. Moreover, the model is also generalized in [1] to a multiple set of leaders; both these generalizations will be used in the present paper.

## 3. Numerical results with first- and second-neighbours interaction

The first set of numerical simulations will concern the behaviour of systems with first- and second-neighbours interaction (respectively FNI and SNI), and without the implementation of the fracture algorithm. As already said, the idea is to show the similarities of the discrete system here considered with deformable bodies characterized by first or second gradient deformation energy densities. In [1] some results in this direction were already presented and various types of external actions have been considered, in particular pulling vertices at 45 degrees with a uniform or accelerated motion.

In the present paper, we study other kinds of external actions, specifically the ones involving shear deformation. In the first set of numerical results, in Figure 3 and Figure 4, the left side has been clamped, while the elements of the right side move in the positive $y$ direction with a velocity of 0.02 $l/t$ respectively for the FNI and SNI case. These results have been compared with analogous continuous systems (Figure 5(a) first gradient and

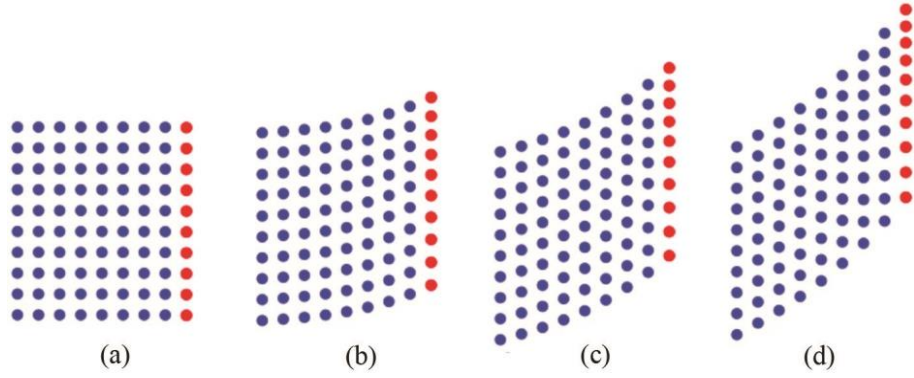

**Figure 3.** FNI system subject to shear external action: the right side (in red) is pulled in the positive $y$ direction, the left side is clamped. Actual configurations for four instants of time are plotted: (a) $t = 1$; (b) $t = 30$; (c) $t = 60$; (d) $t = 90$.

(b) second gradient). For continuum simulations, a standard isotropic two-dimensional first gradient model has been employed, while for the second gradient case we used the general form of isotropic deformation energy:

$$U(G_{ij}, G_{ij,h}) = \frac{\lambda}{2} G_{ii} G_{jj} + \mu G_{ij} G_{ij} + 4\alpha_1 G_{aa,b} G_{bc,c} + \alpha_2 G_{aa,b} G_{cc,b} + 4\alpha_3 G_{ab,a} G_{cb,c} + 2\alpha_4 G_{ab,c} G_{ab,c} + 4\alpha_5 G_{ab,c} G_{ac,b},$$

where $G$ is the nonlinear Green-Saint Venant strain tensor, $\lambda$ and $\mu$ the Lamé coefficients and $\alpha_i$ are second gradient constitutive parameters. In particular in the first gradient case the Lamé coefficients are:

$$\lambda = 150\,[\text{MPa m}], \quad \mu = 2\,[\text{MPa m}],$$

while in the second gradient case we set:

$$\lambda = 150\,[\text{MPa m}], \quad \mu = 150\,[\text{MPa m}], \quad \alpha_1 = 1[El_c^2],$$
$$\alpha_2 = 1[El_c^2], \quad \alpha_3 = 2[El_c^2], \quad \alpha_4 = 1[El_c^2], \quad \alpha_5 = 0.5[El_c^2],$$

where $[E]$ is the generalized Yang modulus and $[l_c]$ a characteristic length (see [21]). A first gradient theory is recovered with $\alpha_1 = 0, ..., \alpha_5 = 0$.

In the deformed configuration, one can see how different convexity in the horizontal sides, found in the continuous case, are recovered in the discrete one, as highlighted in Figures 6 and 7. Some further specific characteristics of the geometry of the side are finely reproduced by the discrete model, as, for instance, the fact that, in the second gradient/SNI case, the curve representing the side has a local maximum and then an inflexion point, while in first gradient/FNI case no such effect is practically visible. On the whole, one can notice that SNI entails a more compact deformed shape and less pronounced convexity of the sides, as observed in second gradient case.

Another example of shear action can be found in the next numerical results. In Figure 8 and Figure 9, two opposite vertices are pulled in opposite horizontal directions with velocities of $\pm 0.02\ l/t$ for FNI and SNI respectively. In this case the reader can notice the different propagation of the effects of the external action inside the system, since for SNI interaction, more nonlocal effects are found. In fact, comparing Figure 8(f) and Figure 9(f), it is clear that the deformation in the second case is highly nonlocal if one considers the spacing between the elements of the horizontal sides of the system.

In some particularly simple cases of imposed external action, less evident differences appear between FNI and SNI systems. For instance, in Figure 10 and Figure 11 the results of a uniaxial traction applied to the right side of the system is shown (the left side is clamped). The leaders are pulled with a velocity of $0.02\ l/t$ in the positive $x$ direction. In this case, no substantial differences in the behaviour of the systems is found between FNI and SNI.

## 4. Fracture

In the presented model, the fracture has been introduced as a loss of interaction between neighbouring elements. In particular, we introduced a threshold value (that will be indicated with ), which is the Euclidean distance evaluated in the actual configurations of the system. When this threshold is overcome, the interaction is broken, that is, the position of the centroid for the evolution of one element does not anymore take into account the element whose distance overcame the threshold. Instead, a new fictitious element is introduced to preserve, as in the case of boundary elements, the symmetry of the Lagrangian neighbours. We will call 'post-fracture' elements the fictitious elements introduced in this way (for a graphical representation see Figure 12).

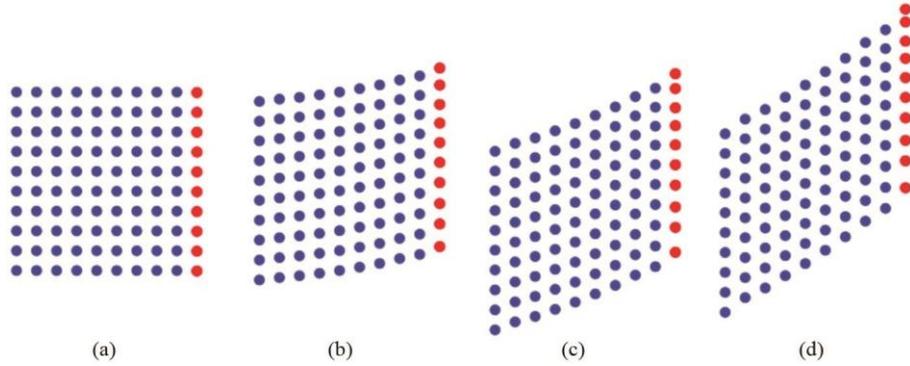

**Figure 4.** SNI system subject to shear external action: the right side (in red) is pulled in the positive $y$ direction, the left side is clamped. Actual configurations for four instants of time are plotted: (a) $t = 1$; (b) $t = 30$; (c) $t = 60$; (d) $t = 90$.

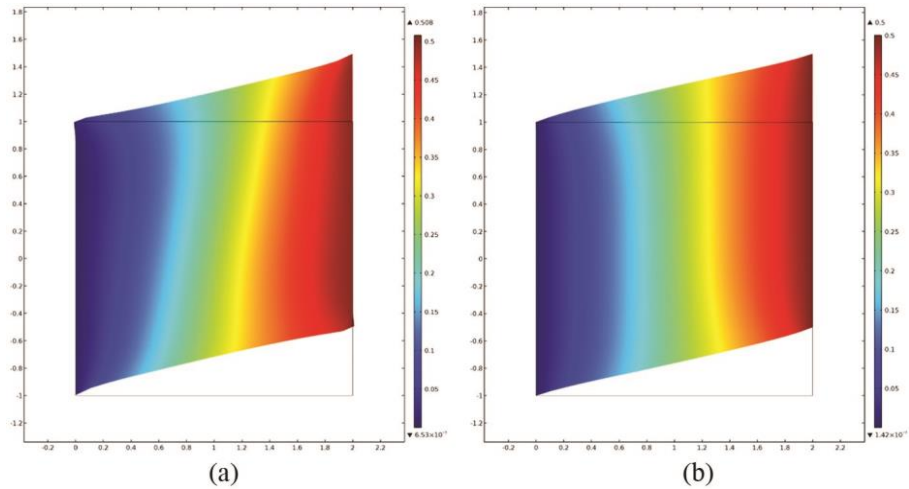

**Figure 5.** Shear continuum simulations: first gradient constitutive relations (a) and second gradient constitutive relations (b).

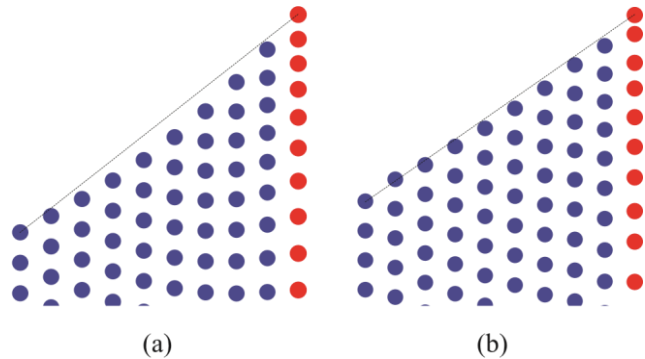

**Figure 6.** Comparison between FNI (on the left) and SNI (on the right) systems subject to shear external action. (a) $t = 90$. (b) $t = 90$.

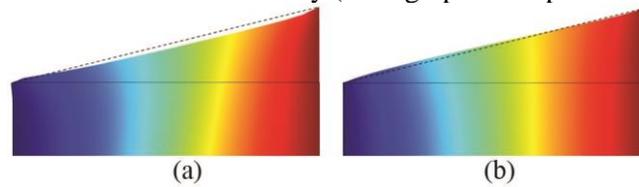

**Figure 7.** Particular of shear continuum simulations: first gradient constitutive relations (a) and second gradient constitutive relations (b).

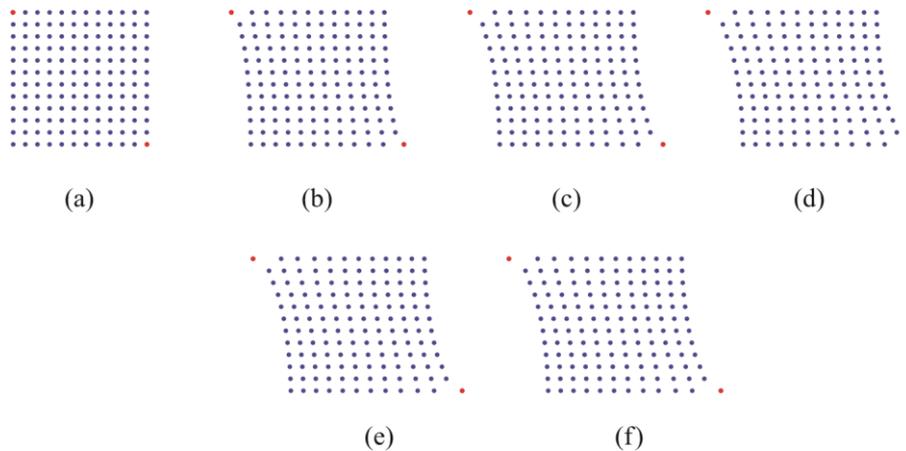

**Figure 8.** FNI interacting system subject the pulling of two opposite vertices in opposite horizontal directions. Actual configurations for six instants of time are plotted: (a) $t = 1$; (b) $t = 100$; (c) $t = 200$; (d) $t = 300$; (e) $t = 400$; (f) $t = 500$.

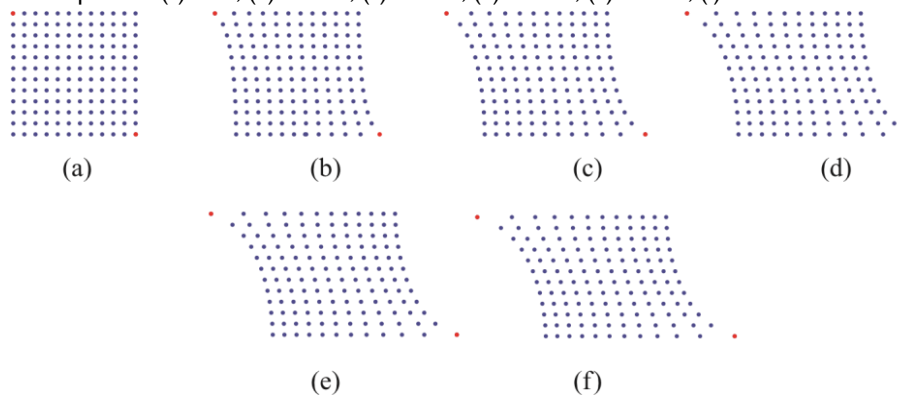

**Figure 9.** SNI interacting system subject the pulling of two opposite vertices in opposite horizontal directions. Actual configurations for six instants of time are plotted: (a) $t = 1$; (b) $t = 100$; (c) $t = 200$; (d) $t = 300$; (e) $t = 400$; (f) $t = 500$.

The information about the interaction between elements in an evolving system is taken into account by means of a three-indices object, that will be referred to as the adjacency matrix, and will be denoted by $A_{ijk}$. The generic element of the adjacency matrix is a boolean variable (TRUE if the interaction is still present, FALSE otherwise); the first two indices uniquely determine an element in the swarm while the third one runs over the possible interactions that this particle can have (so an order in the neighbouring particles has to be introduced). In this way, in the evolution of the system, one can keep track of the interacting elements and introduce postfracture fictitious elements when needed. Moreover, post-fracture fictitious elements play a very important role in the behaviour of the system after fracture. In fact, varying the fixed distances of the new fictitious element (that will be denoted with  in the following) from the true elements, plastic-like and elastic-like post-fracture behaviours can be recovered, where it is intended by 'elastic behaviour' in fracture, the property of the fracture edges, or of the disconnected pieces originated after fracture has occurred, to recover the original shape. The previous algorithm can be easily generalized to SNI by introducing two (possibly different) thresholds for the two orders of neighbours. In this case, two orders of post-fracture fictitious elements (i.e. first neighbour and second neighbour) arise.

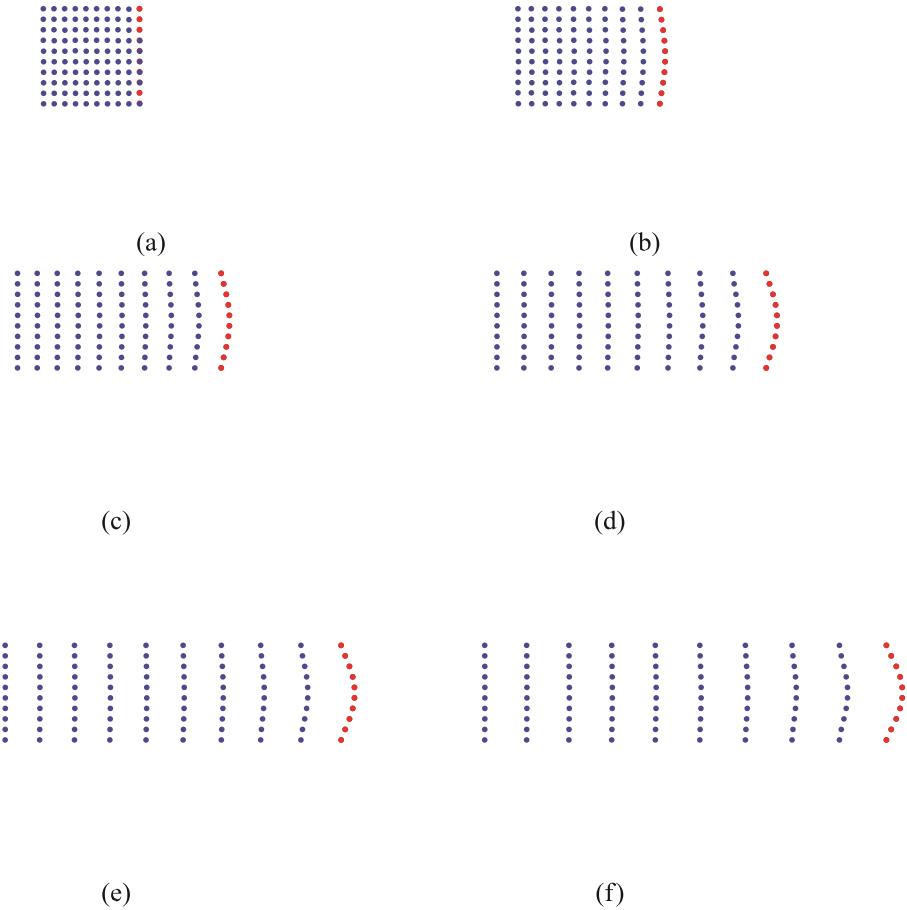

**Figure 10.** FNI system with the left side clamped and the right one pulled horizontally in the positive $x$ direction. Actual configurations for six instants of time are plotted: (a) $t = 1$; (b) $t = 60$; (c) $t = 120$; (d) $t = 180$; (e) $t = 240$; (f) $t = 300$.

In Figure 13 a basic case is presented: two halves of the bottom side of the system are horizontally pulled in opposite directions (the elements move at a velocity of 0.01 $l/t$, =1.8 $l$, and =1 $l$) and the result is a fracture propagating inside the system.

As anticipated, an important role is played by , which is capable of determining different behaviours of the system after fracture has occurred. In Figure 14 and Figure 15 a short parametric study is shown. In the first case a set of −1), Λ =2.5 elements neighbouring the right bottom vertex move with a velocity of 0.03 $l/t$ at 45 degrees in the direction (1,$l$ and =0.93 $l$, while in the second case analogous external action is applied, but it has been set to Λ =2.5 $l$ and =1.3 $l$. As can be seen, for <1 $l$, where the unit is the step of the lattice in the reference configuration, the fracture wedge shrinks, showing a behaviour typical of certain systems displaying surface tension effects. In the second case, we fixed =1.3 $l$, and it is clear that for values of >1 $l$ the system shows plastic behaviour after the fracture, retaining the fracture front shape instead of recovering its original shape.

In Figure 16 three elements belonging to the two horizontal sides and near opposite vertices are pulled in opposite directions with a velocity of ≈0.02 $l/t$ in the directions (−2,−1) and (2,1), =1.8 $l$ and =1 $l$. As one can see, a complex fracture, with a far-reaching crack, arises. The leaders disconnect from the system carrying with them three other elements each. In the next set of simulations it is shown how the geometry of the system and of the imposed external action affect the shape of the fracture front. In Figure 17 two central elements of the bottom side are pulled respectively in the (−1,−1) and (1,−1) directions at a velocity of 0.02 $l/t$ with =1.2 $l$ and = 1. In the next simulation instead we consider an external action applied pulling uniformly the bottom left and right corner of the system respectively in the direction (1,−1) and (−1,−1). The fracture threshold is set to 1.17 $l$, and in this case we also vary the parameter (representing as we recall the distance at which post-fracture fictitious elements arise), which is set to 0.93 $l$. The result, showing a trapezoidal fracture front, is visible in Figure 18. We compare this with a continuum simulation (performed with COMSOL) in which an analogous displacement is imposed to two small regions around the bottom vertices of a squared sample. The result is plotted in Figure 19, where the curves represent level lines of the eigenvalue relative to the compressive principal stress directions. The numerical results indicate that the crack would probably (i.e. in absence of relevant defects) originate close to the regions in which the external action is imposed, with an angle of approximately 45 degrees to the sides of the body, which is in good agreement with the fracture edge observed in the discrete system. The compression level lines, being orthogonal to the maximum traction directions, indicate the probable fracture edge.

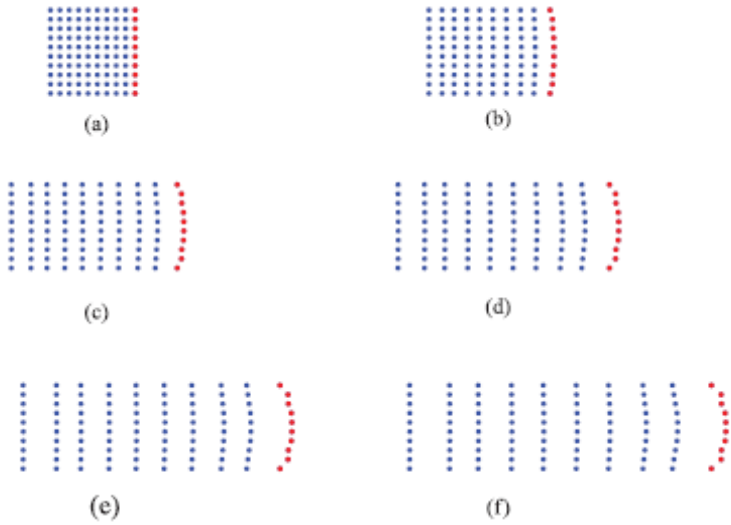

**Figure 11.** SNI system with the left side clamped and the right one pulled horizontally in the positive $x$ direction. Actual configurations for six instants of time are plotted: (a) $t = 1$; (b) $t = 60$; (c) $t = 120$; (d) $t = 180$; (e) $t = 240$; (f) $t = 300$.

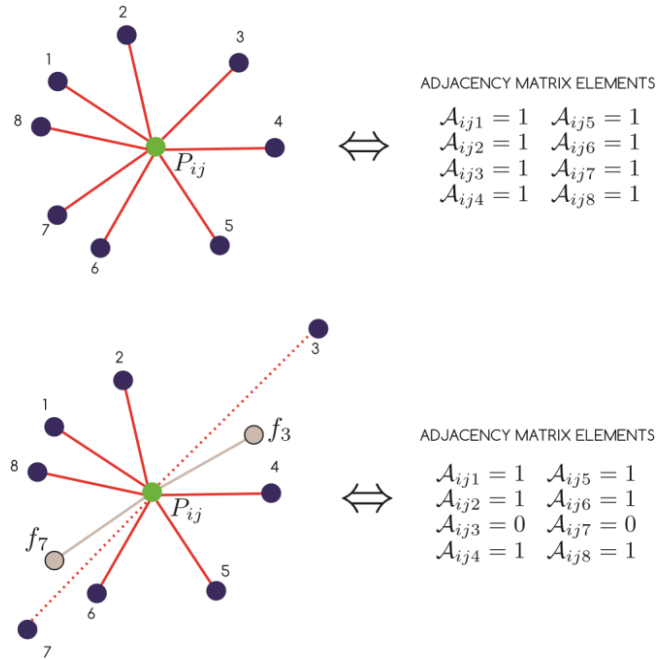

**Figure 12.** Graphical representation of the loss of interaction. Top: the generic element $P_{ij}$ is interacting with all his neighbours, that is, all the elements $A_{ijk}$ for $k \in \{1,...,8\}$ are equal to one. Bottom: the threshold has been overcome by the distance between $P_{ij}$ and the neighbours labelled with 3 and 7, the corresponding elements of the adjacency matrix, that is, $A_{ij3}$ and $A_{ij7}$ are set to 0, and the fictitious elements $f_3$ and $f_7$ enter the computation of the centroid.

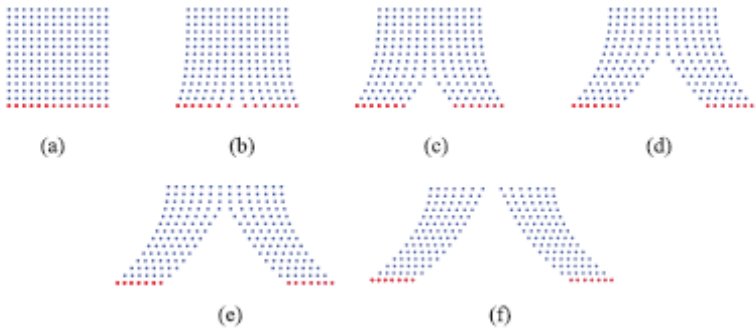

**Figure 13.** System subject to the pulling of two halves of the bottom side in opposite directions. Actual configurations for six instants of time are plotted: (a) $t = 1$; (b) $t = 48$; (c) $t = 96$; (d) $t = 144$; (e) $t = 192$; (f) $t = 240$.

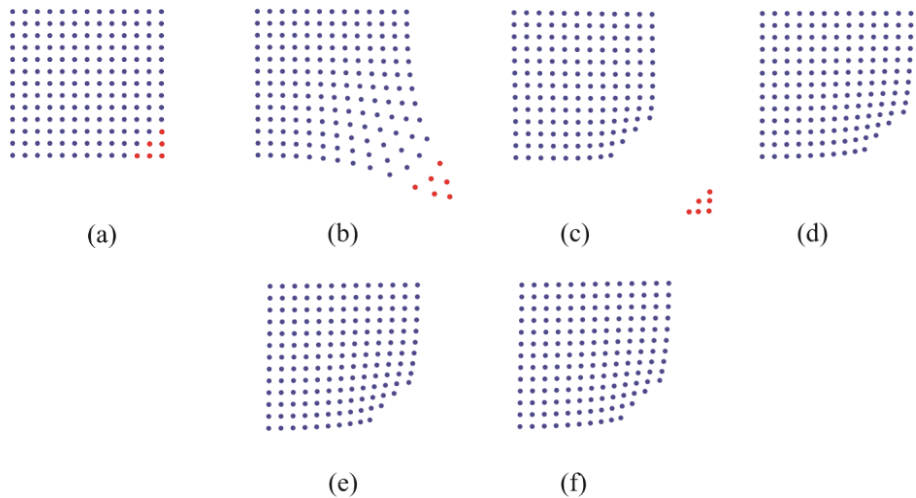

**Figure 14.** Parametric study of fracture: in this figure the behaviour of the system for < 1$l$ is shown. Actual configurations for six instants of time are plotted: (a) $t = 1$; (b) $t = 30$; (c) $t = 60$; (d) $t = 90$; (e) $t = 120$; (f) $t = 150$.

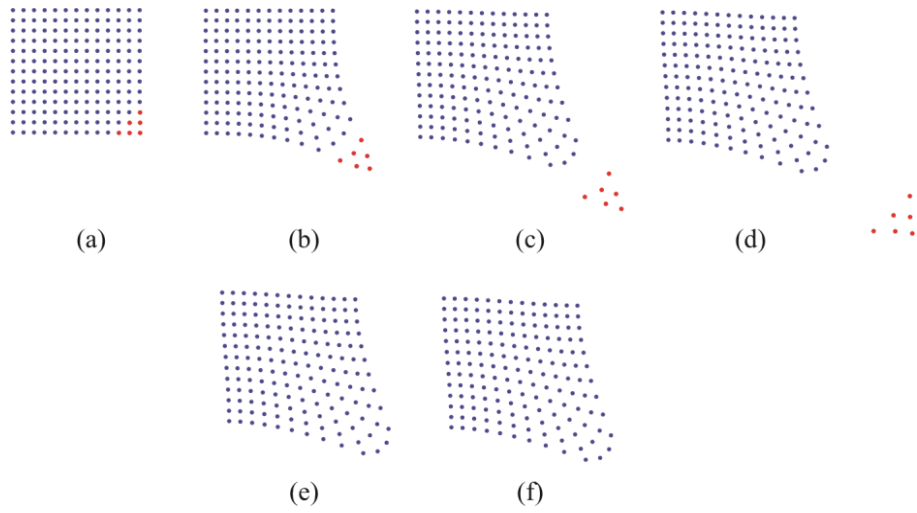

**Figure 15.** Parametric study of fracture: in this figure the behaviour of the system for > 1l is shown. Actual configurations for six instants of time are plotted: (a) $t = 1$; (b) $t = 30$; (c) $t = 60$; (d) $t = 90$; (e) $t = 120$; (f) $t = 150$.

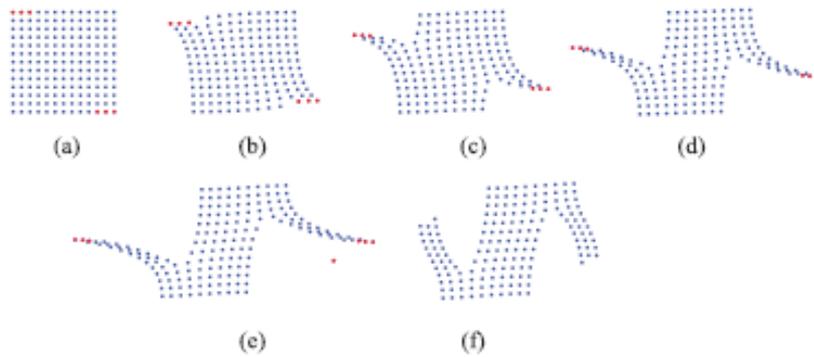

**Figure 16.** System subject to the pulling of three elements in the top side and three elements in the bottom side in opposite directions. Actual configurations for six instants of time are plotted: (a) $t = 1$; (b) $t = 60$; (c) $t = 120$; (d) $t = 180$; (e) $t = 240$; (f) $t = 300$.

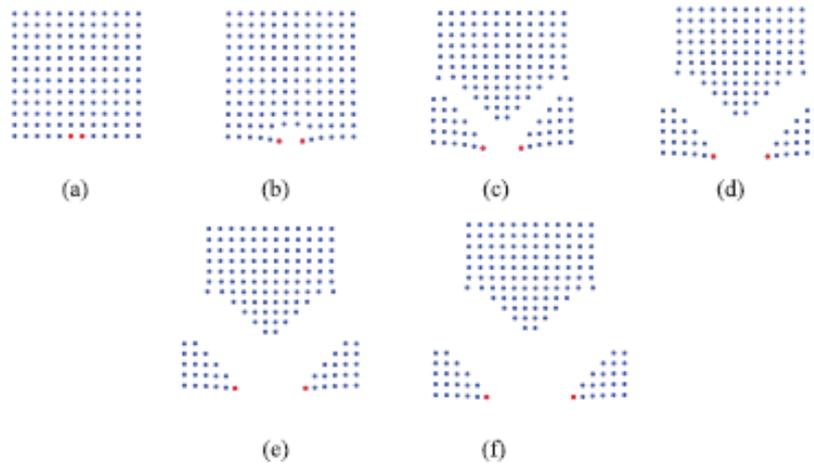

**Figure 17.** System subject to the pulling of two central elements of the bottom side. The left leader is pulled in the $(-1, -1)$ direction, the right one in the $(1, -1)$. Actual configurations for six instants of time are plotted: (a) $t = 1$; (b) $t = 40$; (c) $t = 80$; (d) $t = 120$; (e) $t = 160$; (f) $t = 200$.

As anticipated, the proposed model of fracture can be easily generalized to $n$th neighbour interactions. In Figure 20, in fact, we show an SNI system where the left bottom vertex is pulled in the direction $(1,-1)$ while the right one is pulled in the direction $(-1,-1)$. In this case two thresholds have to be introduced for first and second neighbours, and also two values, and , are needed to account for the distance at which the post-fracture fictitious elements arise. In the case in consideration, we chose $\Lambda_1 = 0.4$, $\Lambda_2 = 1.8$, $\Gamma_1 = 1.05$, $\Gamma_2 = 2.05$. An interesting feature of SNI fracture is the possibility of modelling periodic crack formation (which is well known in some micro-structured materials as for instance concrete), and indeed this kind of behaviour is visible in Figure 20. The onset of this kind of behaviour with such a simple model represents, in the opinion of the authors, a promising direction to be exploited in future investigations. Finally, we remark that the external action is the same as in the simulation of Figure 18, which shows how deeply the switch from FNI to SNI can affect the overall behaviour of the system.

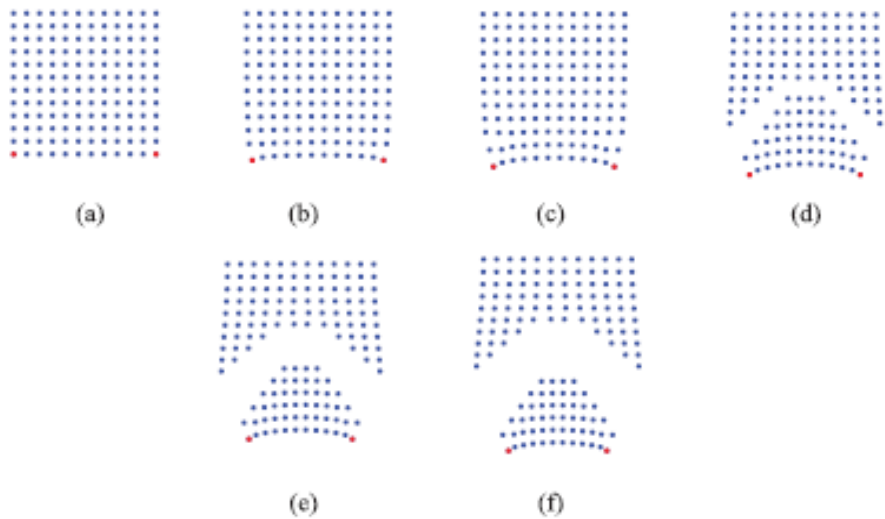

**Figure 18.** Evolution of fracture for an FNI interacting system with two vertex pulled in $(1,-1)$ and $(-1,-1)$ directions. Actual configurations for six instants of time are plotted: (a) $t = 1$; (b) $t = 40$; (c) $t = 80$; (d) $t = 120$; (e) $t = 160$; (f) $t = 200$.

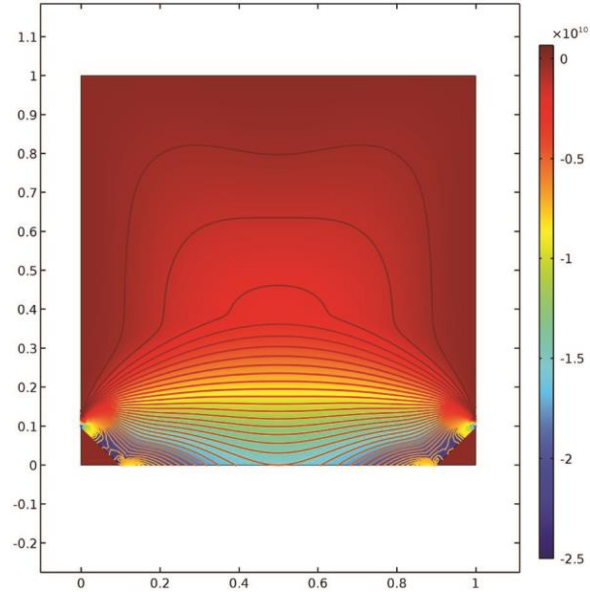

**Figure 19.** Level lines of compressive principal stress direction (and the relative eigenvalue in the color map) in the case of continuum finite element simulation with an imposed displacement similar to the one considered in the previous discrete simulation.

## 5. Conclusions

In this work we showed some cases of numerical simulations concerning a discrete mechanical system presented in [1]. The system is characterized by a centroid-based interaction law and an interaction ranging from first to second neighbours. The results showed good similarity between the prediction of the discrete model and the predictions of standard FEM simulations. In particular, some aspects in the geometry of the deformed configurations in the continuous case are found for the discrete system as well. Moreover, a fracture algorithm was introduced and some results, including periodic crack formation, were provided. Also in this case, a comparison with the continuous case shows a good agreement between the fracture front geometry (discrete system) and the most stressed area (continuous system). Being based on a linear operation (the computation of the centroid), the algorithm described in Section 2 entails computational costs that increase only linearly with the number of interacting elements. This is advantageous when compared with the over-exponential growth of computational cost usually associated with FEMs. Despite the interest of the presented results, the model is still at a preliminary stage and a 3D generalization would be very interesting in order to address fracture propagation in real-world bodies. Moreover, a generalization of the interaction algorithm, to encompass the richness of behaviour of different materials (e.g. metallic or plastic) is under study. In fact, a suitable potential interaction could take into account different deformation regimes, such as elastic and plastic ones together with fracture, as shown in this paper.

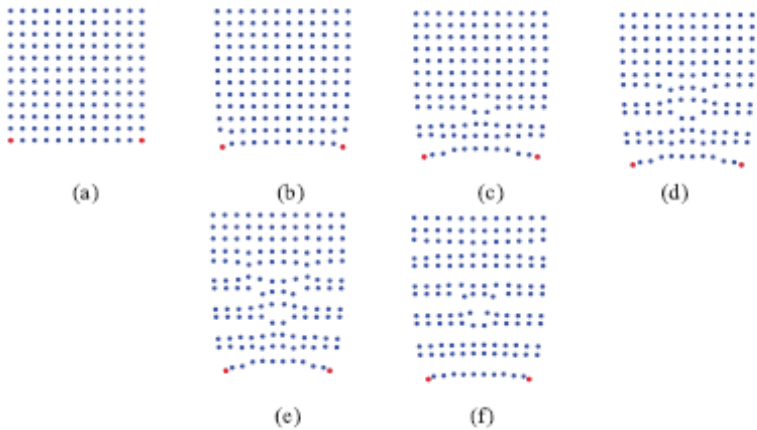

**Figure 20.** Evolution of fracture for an SNI interacting system with two vertex pulled in $(1,-1)$ and $(-1,-1)$ directions. Actual configurations for six instants of time are plotted: (a) $t = 1$; (b) $t = 40$; (c) $t = 80$; (d) $t = 120$; (e) $t = 160$; (f) $t = 200$.

Finally, some issues still concern the algorithm. When one acts on only one leader, the external action can be considered as a 'true' imposed motion on the leader. When one acts on multiple leaders, instead, the system evolves by single-leader system superposition, which leads to the fact that leaders actually interact during the evolution. In fact, one should intend the action as an 'imposed external action', more than an 'imposed external motion' (this concept is clear looking at Figures 10 and 11 where the same motion is imposed on all the elements of the side, but the external action is such to give a bell-shaped side in the deformed configurations). This strategy has been adopted mainly for two reasons. The first one is that by the single-leader system superposition the effect due to each leader decreases with the (Lagrangian) Chebyshev distance on the lattice. The second one is that in the case of independent multiple leaders, some elements of the system could interact in different virtual steps with different leaders (the action due to a closer leader affects a certain element in an earlier virtual time than the action due to a less close leader), while in our simulations the action due to all the leaders affects in the same virtual time a generic element, with an intensity depending on the Lagrangian Chebyshev distance between them. Of course, the proposed algorithm could be improved, but in our opinion one should try to keep the aforementioned peculiarities (which are sensible on physical grounds) while achieving independence between the leaders. From the theoretical point of view, it would be very interesting to find an explicit form for the potential interaction, and in fact we speculate that the model, in the pre-fracture regime, is equivalent to one in which particles interact by means of a potential depending on the squared distance, but the explicit form of this potential is still unknown and should be quite complex if one wants to reproduce exactly the combined effect of virtual configurations and fictitious boundary.

## Funding

The authors received no financial support for the research, authorship, and/or publication of this article.

## Note

1. The Chebyshev distance in $\mathbb{R}^2$ is the distance given by $\rho((x^1, x^2), (y^1, y^2)) = \max\left\{|x^1 - y^1|, |x^2 - y^2|\right\}$.